\documentclass[12pt]{article}
\usepackage {amssymb}
\usepackage {amsfonts}
\usepackage {amsmath}
\usepackage {amsthm}
\usepackage {graphicx}
\usepackage {geometry}
\usepackage{fancyhdr}

\newcommand{\p}{\partial}
\newcommand{\dr}{\text{d}}

\newcommand{\ddo}{\ddot{\text{o}}}

\theoremstyle{definition}

\theoremstyle{definition}

\theoremstyle{remark}

\theoremstyle{remark}

\newcounter{sect}
\newenvironment{sect}
    {
    \begin{center}\stepcounter{sect}{\Large \S\thesect}
    }
    {
    \end{center}
    }
    
\newcounter{app}

\newcounter{eq}[sect]
\newenvironment{eq}
	{
	\stepcounter{eq}\begin{equation}
	}
	{
	\tag{\thesect.\theeq}\end{equation}
	}
	
\newcounter{eqa}[app]

\newenvironment{eqa*}
	{
	\stepcounter{eqa}\begin{equation*}\begin{aligned}
	}
	{
	\end{aligned}\tag{A\theapp.\theeqa}\end{equation*}
	}
	
\newenvironment{eq*}
	{
	\stepcounter{eq}\begin{equation*}\begin{aligned}
	}
	{
	\end{aligned}\tag{\thesect.\theeq}\end{equation*}
	}

\begin{document}
\begin{center}
D. D. H. Yee\footnote[1]{Department of Physcis and Astronomy, Hofstra University, Hempstead, New York U.S.}$^,$\footnote[2]{Department of Engineering/Physics/Technology Nassau Community College, Garden City, New York, U.S.}$^,$\renewcommand{\thefootnote}{$\dag$}\footnote{Corresponding author's E-mail: Daniel.D.Yee@hofstra.edu}

Richard Myers$^1$
\end{center}

\vfill

\begin{sect}
{\Large Abstract}
\end{sect}

Landau was the first to advance hydrodynamic concepts such as density and velocity to describe the superfluidity of liquid He$^4$. Due to the recent spectacular success of experiments demonstrating Bose Einstein condensation in dilute Bose atomic gases, interest has been revitalized in the theoretical description of Bose Einstein condensates. Many of the properties of these gases were obtained by using the Gross-Pitaevskii equation (GPE) to derive the hydrodynamic equations for the gases. However, it is interesting to apply the hydrodynamic equations obtained by Yee for bosons. Many of the properties obtained for the dilute Bose gases are also consequences of Yee's hydrodynamic equations, which derive from a formalism distinct from that of the GPE.

\newpage

\begin{sect}
{\Large Introduction}
\end{sect}

Soon after the discovery of liquid helium's superfluidity by Kapitza \cite{kapitza}, and independently, Allen and Misener \cite{allen_misener}, Landau\cite{landau} advanced his celebrated theory of superfluidity to describe liquid Helium's properties by quantizing the phenomological classical hydrodynamic equations. The dynamical variables in the classical theory were the density $\rho$ and the velocity field $\bold v$. In Landau's theory, these variables were then promoted to operators whose commutation relations he determined by summing over the typical position and momentum operators of quantum theory. In terms of these density and velocity operators, he used a phenomenological Hamiltonian to successfully treat the dynamics of superfluid liquid Helium. Interestingly enough, he did not invoke the concept of Bose condensation as advocated by L. Tisza \cite{tisza} and London \cite{london}, whose position was later supported by Feynman \cite{feynman}. To account for superfluidity, Landau proposed that the lowest excited states were the collective excitations of the liquid called phonons. Subsequently, quantization of the classical hydrodynamic variables was treated by Kronig and Thellung \cite{kronig_thellung,thellung}. However, it is important to observe that this and similar theories were not derived from the first principles of the established theory of microscopic particles.

During the 60's and 70's, advances were made in the use of current algebras to describe many properties of hadron physics. During this time, Dashen and Sharp \cite{dashen_sharp} published their classic paper discussing the possibility of formulating relativistic field theories in terms of currents, that is, in treating these currents as the fundamental variables of the system rather than the underlying canonical fields. In particular, they pointed out that the use of currents had long been established in classical non-relativistic field theories such as hydrodynamics. Since the fundamental variables in hydrodynamics were density and velocity, as advocated by Landau, rather than density and current, Yee \cite{yee} extended the hydrodynamic current approach to density and velocity fields so that many of the concepts and interpretations of classical hydrodynamic theory might be exploited to guide and interpret the results and developments of the quantum theory as Landau did in developing his theory of superfluidity.

From the density and current commutation relations which derive from first principles in quantum field theory, Yee determined the density and velocity commutation relations. In the process of this derivation Yee rediscovered the quantum version of the Clebsch transformation \cite{clebsch,ziman} which has played an essential role in the more recent Hamiltonian formulation of classical hydrodynamics. Consequently, Yee derived a system of hydrodynamic equations for Bose systems.

In the last three decades, a remarkable series of experiments \cite{exp1,exp2} involving dilute Bose gases has revitalized interest in quantum gases and their theoretical descriptions\renewcommand{\thefootnote}{$1$}\footnote{For a modern summary of theoretical results, the reader is referred to \cite{book}.}, particularly due to their confirmation of the theoretical framework developed by Bogoliubov, Lee, Huang, Yang and many others \cite{group}. Indeed, most recent theoretical treatments base their efforts on the celebrated Gross-Pitaevskii equation (GPE) \cite{gross,pitaevskii} as an extension of Bogoliubov's framework to non-uniform trapping potentials. The GPE comes from the first quantization scheme, utilizing the Hartree-Fock approximation together with an s-wave pseudopotential to modify the many boson Sch$\ddo$dinger equation. However, Ruckenstein in 2001 \cite{ruckenstein} has revisited the current algebra formalism of Dashen, Sharp, Yee, and others to describe the Bose gas. In this paper we continue in the formalism of Yee to show correspondence to and extension of some common results from the GPE approach as well as the approach presented by Ruckenstein, which uses density and current rather than the density and velocity of Yee as the dynamical variables. Furthermore, the nature of the second quantization scheme allows for a simple and clear approximation to Bose condensates with the c-number limit described in \S3 and removes the need to assume an s-wave pseudopotential, though we will still make this assumption to better show correspondence of results.

\vspace{.2 in}

\begin{sect}
{\Large Velocity Operator Approach}
\end{sect}

The formalism presented by Yee \cite{yee} begins from first principles in the second quantization formalism with the typical density and current operators
\begin{eq*}
&\rho(\bold x)=m\psi^\dag(\bold x)\psi(\bold x),\\
&\bold J(\bold x)=\frac{\hbar}{2i}\left[\psi^\dag(\bold x)\nabla\psi(\bold x)-\nabla\psi^\dag(\bold x)\psi(\bold x)\right],
\end{eq*}
where $m$ is the mass of a particle in our gas. These definitions then imply the commutation relations
\begin{eq*}
&\left[\rho(\bold x),\rho(\bold y)\right]=0,\\
&\left[\rho(\bold x),J_j(\bold y)\right]=-i\hbar\frac{\p}{\p x_j}\left[\delta(\bold x-\bold y)\rho(\bold x)\right],\\
&\left[J_k(\bold x),J_j(\bold y)\right]=-i\hbar\frac{\p}{\p x_j}\left[\delta(\bold x-\bold y)J_k(\bold x)\right]+i\hbar\frac{\p}{\p y_k}\left[\delta(\bold x-\bold y)J_j(\bold y)\right],
\end{eq*}
which we will henceforth take as the defining relations of $\rho$ and $\bold J$, viewing (3.1) as no more than a concrete realization of the operators which are defined to satisfy (3.2). Furthermore, we will observe that the many body extension of (3.1) also satisfies (3.2), making the results of this formalism applicable to many-body problems.

Now, as discussed by Yee, there are difficulties associated with defining the velocity operator, however, we may take as definition the form
\begin{eq}
\bold v(\bold x)=\frac{1}{2}\left(\frac{1}{\rho(\bold x)}\bold J(\bold x)+\bold J(\bold x)\frac{1}{\rho(\bold x)}\right),
\end{eq}
which was originally suggested by Landau to be the second quantized analogue of the classical velocity.

From the commutation relations (3.2) which define the current algebra, we may establish the commutation relations between the velocity operator and the canonical variables, $\rho$ and $\bold J$. The determination of these relations, particularly the velocity-velocity relation, is presented by Yee in \cite{yee} and result the quantized version of the Clebsch formula \cite{clebsch} for the velocity operator. That is,
\begin{eq}
\bold v(\bold x)=-\nabla\phi(\bold x)-\frac{1}{2\rho(\bold x)}\left[\sigma(\bold x)\nabla\lambda(\bold x)+\nabla\lambda(\bold x)\sigma(\bold x)\right].
\end{eq}
This formulation encapsulates both rotational and irrotational flow.

The standard Hamiltonian describing a system of identical particles may now be written in terms of the current and density operators to find
\begin{eq}
H=\frac{1}{8}\int\dr^3x\left[\nabla\rho(\bold x)-2i\bold J(\bold x)\right]\frac{1}{\rho(\bold x)}\left[\nabla\rho(\bold x)+2i\bold J(\bold x)\right]+H_\text{int},
\end{eq}
where $H_\text{int}$ is the interaction term of the Hamiltonian,
\begin{eq}
H_\text{int}=\frac{1}{2}\int\dr^3x\dr^3yV(|\bold x-\bold y|)\rho(\bold x)\rho(y)+\int\dr^3xV_\text{ext}(\bold x)\rho(\bold x).
\end{eq}
This is the Hamiltonian obtained by \cite{dashen_sharp} which may then be rewritten in terms of the operators $\phi,\sigma$ and $\lambda$ to obtain the equations of motion which determine the system dynamics. The complete form of this system is presented in \cite{yee}. For our purposes, it will be sufficient to present the c-number function limit of these equation. This limit, first suggested by Bogoliubov, corresponds to a restriction of our attention to a single state as is the case for Bose gases. We shall abuse notation to avoid introducing new notations for the eigenfunctions of the operators determining this system. So, we write
\begin{eq*}
&\frac{\p\rho(\bold x)}{\p t}+\nabla\cdot\left(\rho(\bold x)\bold v(\bold x)\right)=0,\\
&\frac{\p\lambda(\bold x)}{\p t}+\bold v(\bold x)\cdot\nabla\lambda(\bold x)=0,\\
&\frac{\p\sigma(\bold x)}{\p t}+\nabla\cdot\left(\sigma(\bold x)\bold v(\bold x)\right)=0,\\
&\frac{\p\phi(\bold x)}{\p t}=\frac{\left[\nabla\rho(\bold x)\right]^2}{8\rho^2(\bold x)}-\frac{\nabla^2\rho(\bold x)}{4\rho(\bold x)}-\bold v(\bold x)\cdot\nabla\phi(\bold x)-\frac{1}{2}\bold v(\bold x)\cdot\bold v(\bold x)+k,
\end{eq*}
where $k=\int\dr^2yV(|\bold x-\bold y|)\rho(\bold y)$ and we have, for the time being, neglected the external potential. In particular, we observe that the first and third equations represent continuity equations while the second is equivalent to conservation of $\lambda$ and the fourth represents the equation of motion for the system.

\vspace{.2 in}

\begin{sect}
{\Large Density Fluctuation Expansion}
\end{sect}

We seek to find an equation for the time dependent density fluctuations above the ground state. To begin, we will suppose that the system energies are sufficiently low that excitations are small and so that the system (3.7) is a good approximation to the full operator equation presented in Yee \cite{yee}. We will further only consider irrotational flow for simplicity of our results. Thus, we may take $\lambda=0$ and $\sigma=0$, reducing (3.7) to only the first and final equations:
\begin{eq*}
&\frac{\p\rho(\bold x)}{\p t}=\nabla\cdot\left(\rho(\bold x)\nabla\phi(\bold x)\right),\\
&\frac{\p\phi(\bold x)}{\p t}=\frac{\left[\nabla\rho(\bold x)\right]^2}{8\rho^2(\bold x)}-\frac{\nabla^2\rho(\bold x)}{4\rho(\bold x)}+\frac{1}{2}\left(\nabla\phi(\bold x)\right)^2+k.
\end{eq*}

Since the excitations from the ground state are assumed to be small, we shall expand the density as $\rho(\bold x,t)=\rho_0(\bold x)+\eta(\bold x,t)$ where $\rho_0$ is the ground state density function. This expansion of $\rho$ then requires that $\rho_0$ and $\eta$ must satisfy
\begin{eq*}
&\int\dr^3x\rho_0(\bold x)=mN,\\
&\int\dr^3x\eta(\bold x)=0.
\end{eq*}
Taking a time derivative of the continuity equation and expanding to first order in $\eta$, we find
\begin{eq}
\frac{\p^2\eta}{\p t^2}=\left(\nabla^2\phi+\nabla\phi\cdot\nabla\right)\frac{\p\eta}{\p t}+\left(\nabla\rho\cdot\nabla+\rho\nabla^2\right)\frac{\p\phi}{\p t}.
\end{eq}
But the time derivatives of $\eta$ and $\phi$ may be removed by application of equations (4.1), the second of which expands to
\begin{eq*}
\frac{\p\phi}{\p t}=&\frac{(\nabla\eta)^2}{8\rho_0^2}-\frac{1}{4\rho_0}\left(1-\frac{\eta}{\rho_0}\right)\nabla^2\eta+\frac{1}{2}(\nabla\phi)^2+k-\frac{1}{4\rho_0}\left(1-\frac{\eta}{\rho_0}+\frac{\eta^2}{\rho_0^2}\right)\nabla^2\rho_0\\
&+\frac{1}{8\rho_0^2}\left(1-\frac{\eta}{\rho_0}+\frac{\eta^2}{\rho_0^2}\right)(\nabla\rho_0)^2+\frac{1}{4\rho_0^2}\left(1-\frac{\eta}{\rho_0}\right)\nabla\eta\cdot\nabla\rho_0.
\end{eq*}
We shall additionally assume that the interacting potential between the particles is characterized by the s-wave pseudopotential, $V(|\bold x-\bold y|)=g\delta(\bold x-\bold y)$, where $g=\frac{4\pi\hbar^2a}{m}$ and $a$ is the s-wave scattering length. This then allows us to write $k=g\rho$ so the expansion of (4.3) may be written in the form
\begin{eq}
\frac{\p^2\eta}{\p t^2}=g\nabla\cdot[\rho_0\nabla\eta]+\kappa,
\end{eq}
where $\kappa$ is given by
\begin{eq*}
\kappa=\nabla^2&\phi\nabla\phi\cdot\nabla\eta+\nabla\cdot\left[\nabla\phi\left([\rho_0\nabla\phi]+\eta\nabla^2\phi\right)\right]+\left(\nabla\cdot[\eta\nabla]\right)\left[\frac{1}{2}(\nabla\phi)^2+g\rho_0\right]\\
&+\nabla\cdot[\rho_0\nabla]\left[\frac{1}{2}(\nabla\phi)^2-\frac{\nabla^2\eta}{4\rho_0}\right]+\frac{g}{2}\nabla^2(\rho_0)^2+\bold D_{\bold v}^2\eta,
\end{eq*}
and $\bold D_{\bold v}=\bold v\cdot\nabla$ is the directional derivative along the flow lines.

\vspace{.1 in}

We now seek to specify the ground state density function by minimizing $H_\text{int}$. The interaction Hamiltonian may be rewritten as
\begin{eq*}
H_\text{int}&=\frac{1}{2}\int\dr^3x\dr^3yV(|\bold x-\bold y|)\rho_0(\bold x)\rho_0(\bold y)+\int\dr^3V_\text{ext}(\bold x)\rho_0(\bold x)\\
&=\frac{1}{2}\int\dr^3xg\rho_0^2(\bold x)+\int\dr^3xV_\text{ext}\rho_0(\bold x)
\end{eq*}
in anticipation for the minimization. From here, we take the usual course of action and introduce a chemical potential to modify the Hamiltonian by $H_\text{int}^\prime=H_\text{int}-\mu mN$. To minimize the system energy, we will when write
\begin{eq}
\frac{\delta H^\prime}{\delta\rho_0(\bold w)}=\frac{\delta}{\delta\rho_0(\bold w)}\left[\frac{1}{2}\int\dr^3xg\rho_0^2(\bold x)+\int\dr^3xV_\text{ext}\rho_0(\bold x)-\mu\int\dr^3x\rho_0(\bold x)\right].
\end{eq}
Thus, if we specialize to the spherical harmonic trap potential, $V_\text{ext}=\frac{1}{2}m\omega_0^2|\bold x|^2$, we find
\begin{eq}
g\rho_0(\bold x)+\frac{1}{2}m\omega_0^2|\bold x|^2-\mu=0.
\end{eq}
The boundary of our fluid in the ground state will then be spherical with radius $R$. Since $\mu$ is a constant, we may evaluate (4.9) on this boundary to observe that $\mu=\frac{1}{2}m\omega^2_0R^2$. Therefore,
\begin{eq}
R=\left(\frac{2\mu}{m\omega_0^2}\right)^{1/2}.
\end{eq}
Next, if we evaluate the integrals
\begin{eq*}
&\int\mu\dr^3x=\frac{4}{3}\pi\mu\left(\frac{2\mu}{m\omega_0^2}\right)^{3/2},\\
&\int V_\text{ext}\dr^3x=\frac{2}{5}\pi m\omega_0^2\left(\frac{2\mu}{m\omega^2_0}\right)^{5/2},
\end{eq*}
we may then impose the normalization condition2 (4.2) on (4.9) to find
\begin{eq}
gN=(a-b)\mu^{5/2}
\end{eq}
where
\begin{eq}
a=\frac{4}{3}\pi\left(\frac{2}{m\omega_0^2}\right)^{3/2},\ \ \ \ b=\frac{2}{5}\pi m\omega_0^2\left(\frac{2}{m\omega_0^2}\right)^{5/2}.
\end{eq}
Recalling now that $g=4\pi\hbar^2a/m$ and making the definition $a_\text{ho}=\sqrt{\frac{\hbar}{m\omega_0^2}}$, we obtain
\begin{eq}
\mu=\frac{\hbar\omega_0}{2}\left(\frac{15Na}{a_\text{ho}}\right)^{2/5},
\end{eq}
in terms of which we may easily express the fluid radius and ground state density,
\begin{eq}
g\rho_0=\frac{1}{2}m\omega_0^2\left(R^2-|\bold x|^2\right).
\end{eq}

But now, if we return to (4.5), neglect $\kappa$, and use (4.15), we find
\begin{eq}
\frac{\p^2\eta}{\p t^2}=\frac{1}{2}m\omega_0^2\nabla\cdot\left[\left(R^2-|\bold x|^2\right)\nabla\eta\right]
\end{eq}
which is the time dependent equation that was produced by Ruckenstein \cite{ruckenstein} from the formalism of Dashen and Sharp \cite{dashen_sharp} using the Thomas-Fermi limit \cite{TF}. The above equation was first studied by Stringari \cite{stringari} and discussed for a variety of different trap geometries. It then follows that (4.5) represents a generalization of the works which produced (4.16), which required a neglect of every term in $\kappa$ to reproduce it.

\vspace{.2 in}

\begin{sect}
{\Large Dispersion Relations}
\end{sect}

To follow the lead of Salazar \cite{salazar} in their work with the GPE\renewcommand{\thefootnote}{$2$}\footnote{For an overview of Bose-Einstein condensation and the application of the GPE to dilute Bose gas condensates, the reader if referred to \cite{pethick}.}, we shall now suppose that the ground state velocity is zero and that fluctuations away from the ground state are small. That is, $\bold v=\delta\bold v$. In the formalism of Yee which we utilized in the previous section, this now implies that we should only keep to first order in $\phi$ if we are to keep only to first order in $\bold v$. Thus, $\kappa$ is reduced to
\begin{eq*}
\kappa&=g\nabla\rho_0\cdot\nabla\eta+g\eta\nabla^2\rho_0+\frac{g}{2}\nabla^2(\rho_0)^2\\
&-\frac{1}{4}\left(\nabla^2\nabla^2\eta+\nabla(\ln\rho_0)\cdot\nabla\nabla^2\eta+\nabla^2(\ln\rho_0)\nabla^2\eta\right).
\end{eq*}
It then follows that (4.5) has the full form
\begin{eq*}
\frac{\p^2\eta}{\p t^2}&=g\rho_0\nabla^2\eta-\frac{1}{4}\nabla^2\nabla^2\eta+2g\nabla\rho_0\cdot\nabla\eta+g\eta\nabla^2\rho_0+\frac{g}{2}\nabla^2(\rho_0)^2\\
&-\frac{1}{4}\left(\nabla(\ln\rho_0)\cdot\nabla\nabla^2\eta+\nabla^2(\ln\rho_0)\nabla^2\eta\right).
\end{eq*}
We observe that the first two terms of (5.2) precisely replicate a result produced by Salazar from the Gross-Pitaevskii equation. In fact, that result is obtained precisely when we assume that the ground state density function is constant, an assumption Salazar also made in obtaining their result.

If we make the assumption that the ground state density function is constant, then take the Fourier transform, we obtain
\begin{eq}
\omega^2=\mu k^2+\frac{1}{4}k^4
\end{eq}
which is the famous Bogoliubov excitation spectrum formula.

\vspace{.2 in}

\begin{sect}
{\Large Conclusion}
\end{sect}

We have shown that the formalism of Yee, utilizing the density and velocity operators in the second quantization formalism, yield precisely the same result as the second quantization approach utilizing the density and current operators in the Thomas-Fermi limit for the harmonic trap potential. Furthermore, this formalism yields a simple route to the well-known Bogoliubov excitation spectrum formula, along with a generalization to the Fourier transform of the spectrum to allow for the more realistic case of non-constant ground state density.

Yee's formalism has the added benefit of simple interpretation in terms of the classical hydrodynamic concepts. However, because second quantization is not a procedure with a generally unique result, it remains possible that the formalism of Yee differs genuinely in some manner from the density and current formulation or the Gross-Pitaevskii picture. Future avenues of research might involve further exploration of Yee's hydrodynamics equations in other cases to better determine the physical applicability of its consequences. Specifically, it might be interesting to determine precisely what the correction terms to the Bogoliubov excitation spectrum formula are in the case of the spherical harmonic trap potential; a case in which the exact form of $\rho_0$ is both known and simple.

\newpage

\begin{sect}
{\Large Acknowledgements}
\end{sect}

One of the authors, Daniel Yee, would like to thank Professor Levine of the Hofstra Physics Department for his continuous encouragement, ever helpful suggestions, and profound interest in physics. The opportunities made available to Daniel Yee at Nassau Community College by Drs. Capria and O'Dwyer are also gratefully acknowledged.

\vspace{.2 in}


\end{document}